# Analytical Solutions of Schrödinger Equation for the diatomic molecular potentials with any angular momentum

**Huseyin Akcay** [1,a] **and Ramazan Sever** [2,b]

[1] Faculty of Engineering, Başkent University, Baglıca Campus, Ankara, Turkey

[2] Department of Physics, Faculty of Arts and Sciences, Middle East Technical University, 06531 Ankara, Turkey

## Abstract

Analytical solutions of the Schrodinger equation are obtained for some diatomic molecular potentials with any angular momentum. The energy eigenvalues and wave functions are calculated exactly. The asymptotic form of the equation is also considered. Algebraic method is used in the calculations.

[a] E-mail: akcay@baskent.edu.tr

[b] E-mail: sever@metu.edu.tr

## 1. Introduction

Diatomic molecular potential are very important to describe the intramolecular and intermolecular interactions and atomic pair correlations in quantum mechanics. Analytical solutions of the Schrodinger equation (SE) with any angular momentum provides important applications in many fields of physics and chemistry for checking and improving models developed to study quantum mechanical systems and also for improvment of the numerical methods. One can have exact solutions for certain potentials. So far, some useful analytical methods have been developed, such as supersymmetry (SUSY) [1,2], Nikiforov-Uvarov method (NU) [3, 4], Pekeris approximation [5], asymptotic iteration method (AIM) [6, 7,8], variational [9], hyperviria perturbation [10], shifted 1/*N* expansion (SE) and the modified shifted 1/*N* expansion (MSE) [11], exact quantization rule (EQR) [12], perturbative formalism [13, 14, 15], polynomial solution [16, 17] , wave function ansatz method [18, 19] , group theory [20, 21] and path integral [22, 23, 24, 25] to solve the radial Schrödinger equation exactly

In this work we find analytical solutions of SE for a class of diatomic potentials with any angular momentum. For these potentials SE can be transformed into a second order differential equation of a certain parametric form. We use algebraic method for the possible solutions. We apply our formulation to several important diatomic potentials and obtain the energy eigenvalues and the corresponding eigenfunctions. These diatomic potentials include: Generalized Morse Potential [26-28], Mie potential [29-36], Kratzer-Fues Potential [37,38], Coulomb Potential, Pseudoharmonic potential [39,40], Noncetral potential [41-47], Deformed Rosen-Morse Potential [48-53], Generalized Woods-Saxon potential [54-60], Pöschl-Teller potential [61-66].

This paper is organized as follows. In Sect. 2 the formulation is introduced. In Sect. 3, Solution of some potentials are presented. Finally, concluding remarks are given in Sect. 4.

## 2 Formulation of the Approach

For many potentials the SE can be transformed into the following second order parametric differential equation.

$$\frac{d^2\psi}{ds^2} + \frac{(c_1 + c_2 s)}{s(1+c_3 s)}\frac{d\psi}{ds} + \frac{1}{s^2(1+c_3 s)^2}[-\Lambda_1 s^2 + \Lambda_2 s - \Lambda_3]\psi = 0 \qquad (1)$$

where $c_i$ and $\Lambda_i$ are some constants. First we look at the asymptotic form of this equation when s becomes very large. We found that depending on the constant $c_3$ there are two possibilities. When $c_3$ is a nonzero constant the last term vanishes faster than the others and it is neglected in the asymptotic equation. If $c_3$ is zero the last term will be present in the asymptotic form. Let us start with a nonzero $c_3$. In this case an asymptotic analysis of Eq.(1) suggests to consider solution of the following form

$$\Psi(s) = (1+c_3 s)^{-p} s^q y(s) \tag{2}$$

where p and q are some arbitrary constants and $y$ is a new function to be determined. When we insert this into Eq.(1) we obtain the following differential equation for $y$

$$s(1+c_3 s)\frac{d^2 y(s)}{ds^2} + A(s)\frac{dy}{ds} + B(s)y(s) = 0 \tag{3}$$

where

$$A(s) = 2q(1+c_3 s) - 2pc_3 s + c_1 + c_2 s,$$

$$B(s) = q(q-1)\frac{(1+c_3 s)}{s} - 2pqc_s + \frac{p(p+1)c_3^2}{(1+c_3 s)}s$$

$$+ \frac{(c_1 + c_2 s)}{s(1+c_3 s)}[q + (q-p)c_3 s] + \frac{(-\Lambda_1 s^2 + \Lambda_2 s - \Lambda_3)}{s(1+c_3 s)}$$

It is convenient to define a new variable $z = 1 + 2c_3 s$ and write Eq. (3) in terms of this variable. Under this transformation we obtain

$$(1-z^2)^2 \frac{d^2 y}{dz^2} + (1-z^2)[\beta - \alpha + (\alpha + \beta + 2)z]\frac{dy}{dz} + R(z)y = 0 \tag{4a}$$

where

$$\alpha = 2q + c_1 - 1, \quad \beta = -2p - c_1 + \frac{c_2}{c_3} - 1, \quad R(z) = r_1 z^2 + r_2 z + r_3. \tag{4b}$$

The coefficients of the polynomial $R$ are defined as

$$r_1 = q(q-1) - 2pq + p(p+1) + \frac{c_2}{c_3}(q-p) - \frac{\Lambda_1}{c_3^2}, \tag{5a}$$

$$r_2 = 2q(q-1) - 2p(p+1) + 2c_1(q-p) + 2\frac{c_2}{c_3}p + 2\frac{\Lambda_1}{c_3^2} + 2\frac{\Lambda_2}{c_3}, \tag{5b}$$

$$r_3 = q(q-1) + 2pq + p(p+1) + 2c_1(q+p) - \frac{c_2}{c_3}(q+p) - \frac{\Lambda_1}{c_3^2} - 2\frac{\Lambda_2}{c_3} - 4\Lambda_3. \quad (5c)$$

In the appendix we simplify Eq. (4a) by arguing that the coefficients of the polynomial R must satisfy $r_2 = 0$ and $r_1 = -r_3$. In the definitions of these parameters there are two arbitrary constants $p$ and $q$. So with proper choices of these constants one can satisfy these conditions. Let us do this. We start with the condition $r_1 + r_2 + r_3 = 0$ and calculate this sum using Eq. (5). We find the following quadratic equation

$$q^2 - (1-c_1)q - \Lambda_3 = 0 \quad (6)$$

and its roots are

$$q_0 = (\frac{1-c_1}{2}) \pm \sqrt{(\frac{1-c_1}{2})^2 + \Lambda_3} \; . \quad (7)$$

This fixes the arbitrary constant $q$ in terms of the parameters of the equation. Similarly the condition that $r_2$ must vanish leads to a quadratic equation for p which can be written as

$$p^2 - Dp - H = 0 \quad (8)$$

and the roots are

$$p_0 = \frac{D}{2} \pm \sqrt{(\frac{D}{2})^2 + H} \quad (9)$$

where

$$D = \frac{c_2}{c_3} - c_1 - 1, \qquad H = \frac{\Lambda_1}{c_3^2} + \frac{\Lambda_2}{c_3} + \Lambda_3. \quad (10)$$

This gives the second arbitrary constant p in the wave function. Inserting $r_1 = -r_3$ in Eq. (4a) we get $R(z) = r_3(1-z^2)$ and thus Eq.(4a) becomes

$$(1-z^2)\frac{d^2y}{dz^2} + [\beta - \alpha + (\alpha + \beta + 2)z]\frac{dy}{dz} + r_3 y = 0. \quad (11)$$

In the Appendix A, we use the polynomial method for the solution of this differential equation and we find that for an acceptable solution $r_3$ must satisfy the condition $r_3 = n(n+\alpha+\beta+1)$ where n is a positive integer. When we insert this into Eq. (10) it becomes the well known Jacobi's differential equation

$$(1-z^2)\frac{d^2y}{dz^2}+[\beta-\alpha+(\alpha+\beta+2)z]\frac{dy}{dz}+n(n+\alpha+\beta+1)y=0 \qquad (12)$$

and its solution are the Jacobi polynomials $P_n^{\alpha,\beta}(z)$. Thus the wave functions are determined in terms of these polynomials. The condition on $r_3$ is obviously a quantization condition. First using Eq. (5c) we express $r_3$ in terms of the determined parameters. For this, we solve Eq.(6) for $\Lambda_3$ in terms of $q_0$ and $c_1$ then we use Eqs. (8,10) and obtain $\Lambda_2$ in terms of $p_0$, D, $\Lambda_1$ and $\Lambda_3$. When we insert these into Eq. (5c) the relation $r_3 = n(n+\alpha+\beta+1)$ takes the following form

$$(q_0-p_0)^2 + (\frac{c_2}{c_3}+2n-1)(q_0-p_0)+n(n+\frac{c_2}{c_3}-1) = \frac{\Lambda_1}{c_3^2} \qquad (13)$$

where $\alpha$ and $\beta$ are also expressed in terms of $q_0$ and $p_0$ by means of Eq. (4b). This equation gives the energy levels in our formulation.

The second alternative was to have $c_3 = 0$ in Eq. (1). For these problems Eq.(1) becomes

$$\frac{d^2\psi}{ds^2} = \frac{(c_1+c_2s)}{s}\frac{d\psi}{ds}+\frac{1}{s^2}(-\Lambda_1 s^2+\Lambda_2 s-\Lambda_3)\psi = 0 \quad . \qquad (14)$$

An asymptotic analysis of this equation suggests a solution of the following form

$$\psi(s) = \exp(-p_1 s)s^{q_1} y_1(s) \qquad (15)$$

where $p_1$ and $q_1$ are some arbitrary constants and $y_1$ is a function of s. Here we have again taken $s^{q_1} y_1$ instead of $y_1$. Thus an expansion of $y_1$ will start with a constant term. When we insert Eq. (15) into Eq. (14) we get the following differential equation

$$\frac{d^2y_1}{ds^2}+[2\frac{q_1}{s}-2p_1+\frac{(c_1+c_2s)}{s}]\frac{dy_1}{ds}+[\frac{q_1(q_1-1)}{s^2}-2\frac{q_1 p_1}{s}+p_1^2$$
$$+\frac{(c_1+c_2s)}{s}(\frac{q_1}{s}-p_1)+(-\Lambda_1 s^2+\Lambda_2 s-\Lambda_3)]y_1 = 0 \qquad (16)$$

As before we define a new variable $z = (2p_1 - c_2)s$ and write Eq. (16) in terms of this new variable. We obtain

$$z^2 \frac{d^2 y_1}{dz^2} + (k+1-z)z \frac{dy_1}{dz} + A_1(z) y_1 = 0 \tag{17}$$

where

$$k = c_1 + 2q_1 - 1 \tag{18}$$

$$A(z) = \frac{(p_1^2 - c_2 p_1 - \Lambda_1)}{(c_2 - 2p_1)^2} z^2 + \frac{(2q_1 p_1 - c_2 q_1 + c_1 p_1 - \Lambda_2)}{(c_2 - 2p_1)} z$$
$$+ (q_1(q_1 - 1) + c_1 q_1 - \Lambda_3) = \gamma_1 z^2 + \gamma_2 z + \gamma_3 \tag{19}$$

Here $\gamma_1, \gamma_2$ and $\gamma_3$ are defined by the coefficients of $z^2, z^1$ and $z^0$ respectively. When we consider Eq. (18) at z=0 we conclude that $\gamma_3$ must be zero. Remember that $y_1$ is not zero at the origin. Using the definition of $\gamma_3$ we get the following equation

$$q_1(q_1 - 1) + c_1 q_1 - \Lambda_3 = 0. \tag{20}$$

We choose the root

$$q_{10} = \frac{(1-c_1)}{2} + \sqrt{(\frac{1-c_1}{2})^2 + \Lambda_3} \tag{21}$$

as the arbitrary constant $q_1$ in Eq. (15). In the Appendix A, we discuss the polynomial method for the solutions of Eq, (18). We show that acceptable solutions can be obtained if we chose $\gamma_1 = 0$ and $\gamma_2 = n$ where n is a positive integer. The condition $\gamma_1 = 0$ gives a quadratic equation

$$p_1^2 - c_2 p_1 - \Lambda_1 = 0 \tag{22}$$

and we take

$$p_{10} = \frac{c_2}{2} + \sqrt{(\frac{c_2}{2})^2 + \Lambda_1} \tag{23}$$

as the second constant $p_1$ in Eq. (15). The condition $\gamma_2 = n$ leads to the following equation

$$c_1 p_{10} - q_{10}(c_2 - 2p_{10}) - \Lambda_2 = n(c_2 - 2p_{10}). \tag{24}$$

This equation gives the energy spectrum for the systems whose wave functions satisfy Eq. (13). The corresponding wave functions can be obtained from Eq. (18). When we insert the values of the parameters $\gamma_i$ into Eq. (18) we get

$$z\frac{d^2 y_1}{dz^2} + (k+1-z)\frac{dy_1}{dz} + ny_1 = 0. \tag{25}$$

This is the well known Laguerre's associated differential equation its solutions are the associated Laguerre polynomials $L_n^k(z)$. Thus the wave function can be written down using Eq. (1) as

$$\psi(s) = \exp(-p_{10}s) s^{q_{10}} L_n^k([2p_{10} - c_2)s]). \tag{26}$$

In the following section we apply these results for several examples.

## 3. Applications

A) $c_3 = 0$ cases.

Case 1: Generalized Morse Potential

We take the Generalized Morse Potential [26-28] as

$$V(x) = V_1 \exp(-2ax) - V_2 \exp(-ax) \tag{27}$$

and define $s = \sqrt{V_1} \exp(-ax)$ to transform the SE to the following form

$$\frac{d^2\psi}{ds^2} + \frac{1}{s}\frac{d\psi}{ds} + \frac{1}{s^2}[-\Lambda_1 s^2 + \Lambda_2 s + \Lambda_3]\psi = 0 \tag{28}$$

where

$$\Lambda_1 = -\frac{2m}{\hbar^2 a^2}, \quad \Lambda_2 = \frac{2m}{\hbar^2 a^2}\frac{V_2}{\sqrt{V_1}}, \quad \Lambda_3 = 4\varepsilon^2, \quad \varepsilon^2 = -\frac{mE}{2\hbar^2 a^2}. \tag{29}$$

Using Eqs. (1,18,20,22) we calculate the parameters as

$$c_1 = 1,\ c_2 = 0,\ c_3 = 0,\quad q_{01} = \sqrt{\Lambda_3},\ p_{10} = \sqrt{\Lambda_1},\quad k = 2\sqrt{\Lambda_3}. \tag{30}$$

Thus the energy spectrum calculated from Eq. (24) is

$$E_n = -\frac{2a^2\hbar^2}{8m}(2n+1-\frac{\sqrt{2m}V_2}{\hbar a\sqrt{V_1}})^2 \tag{31}$$

and the corresponding wave functions are

$$\psi = s^{2\varepsilon} \exp(-\frac{\sqrt{2m}}{\hbar a}s) L_n^{4\varepsilon}(2\frac{\sqrt{2m}}{\hbar a}s). \tag{32}$$

Case 2: Mie Potential

The Mie potential is given by [29-36]

$$V(r) = V_0[\frac{1}{2}(\frac{a}{r})^2 - \frac{a}{r}]. \tag{33}$$

Using the notation $s = r$ one can write the radial SE as

$$\frac{d^2R}{ds^2} + \frac{2}{s}\frac{dR}{ds} + \frac{(-\Lambda_1 s^2 + \Lambda_2 s - \Lambda_3)}{s^2}R = 0 \tag{34}$$

where

$$\Lambda_1 = -\frac{2mE}{\hbar^2},\quad \Lambda_2 = \frac{2maV_0}{\hbar^2},\quad \Lambda_3 = \frac{2m}{\hbar^2}[\frac{a^2V_0}{2} + \frac{\hbar^2\ell(\ell+1)}{2m}]. \tag{35}$$

Comparing Eq. (34) and Eq. (1) we find $c_1 = 2,\ c_2 = 0,\ c_3 = 0$ and using these with Eq. (17) and Eq. (23) we determine $q_{10}$ and $p_{10}$ as

$$q_{10} = \frac{1}{2}(-1+\sqrt{1+4\Lambda_3}),\quad p_{10} = \sqrt{\Lambda_1}. \tag{36}$$

When we insert these into Eq. (24) we get the following energy eigenvalues

$$\sqrt{\Lambda_1}(2n+1+\sqrt{1+4\Lambda_3}) = \Lambda_2 \tag{37}$$

or writing $\Lambda_1$ explicitly we get

$$E_n = -\frac{\hbar^2}{2m}\Lambda_2^2[2n+1+\sqrt{1+4\Lambda_3}]^{-2}. \tag{38}$$

The corresponding wave functions can be written down from Eq. (26) as

$$\psi = s^{\frac{1}{2}(-1+\sqrt{1+4\Lambda_3})}\exp(-i\varepsilon s)L_n^{\sqrt{1+4\Lambda_3}-1}(2i\varepsilon s) \tag{39}$$

where

$$\varepsilon^2 = \frac{2mE}{\hbar^2}. \tag{40}$$

Case 3: Kratzer-Fues Potential

The Kratzer-Fues Potential [37,38] is given is by

$$V(r) = D_e\left(\frac{r-r_e}{r}\right)^2 \tag{41}$$

and the radial SE with this potential can be written as

$$\frac{d^2R_{nl}}{dr^2} + \frac{2}{r}\frac{dR_{nl}}{dr} + \frac{1}{r^2}\left(\frac{2m}{\hbar^2}\right)[(E_{nl}-D_e)r^2 + 2D_e r_e r - (D_e r_e^2 + \frac{\hbar^2 \ell(\ell+1)}{2m})]R_{nl} = 0. \tag{42}$$

By defining s=r this equation can be expressed as

$$\frac{d^2R_{nl}}{ds^2} + \frac{2}{s}\frac{dR_{nl}}{ds} + \frac{1}{s^2}[-\Lambda_1 s^2 + \Lambda_2 s - \Lambda_3]R_{nl} = 0 \tag{43}$$

where

$$\Lambda_1 = \frac{2m}{\hbar^2}(D_e - E_{nl}) = -\varepsilon_{nl}^2, \quad \Lambda_2 = \frac{4mD_e r_e}{\hbar^2}, \quad \Lambda_3 = \frac{2m}{\hbar^2}[D_e r_e^2 + \frac{\hbar^2 \ell(\ell+1)}{2m}]. \quad (44)$$

The parameters are obtained by using Eqs. (1,18,21,23) as follows
$c_1 = 2$, $c_2 = 0$, $c_3 = 0$, $q_{10} = (1/2)(1+\sqrt{1+4\Lambda_3})$, $p_{10} = \sqrt{\Lambda_1}$. Using these in Eq. (24) we get the energy spectrum

$$\varepsilon_{nl} = -\frac{\Lambda_2^2}{[2n+1+\sqrt{1+4\Lambda_2}]^2}. \quad (45)$$

Eq. (26) gives the following formula for the wave functions

$$R_{nl} = A v^{\frac{1}{2}(1+\sqrt{1+\Lambda_3})} \exp(-v/2) L_n^k(v) \quad (46)$$

where $k = 2+\sqrt{1+4\Lambda_3}$ and $v = 2i\varepsilon s$

Case 4: Coulomb Potential

We take $V(r) = -e^2/r$ for this potential and write the radial part of the SE as

$$\frac{d^2 R}{dr^2} + \frac{2}{r}\frac{dR}{dr} + \frac{1}{r^2}[-\Lambda_1 r^2 + \Lambda_2 r - \Lambda_3)]R = 0 \quad (47)$$

where

$$\Lambda_1 = -\frac{2mE}{\hbar^2}, \quad \Lambda_2 = \frac{2me^2}{\hbar^2}, \quad \Lambda_3 = \ell(\ell+1). \quad (48)$$

The following parameters are obtained by using Eqs. (1,18,21,23)

$$c_1 = 2, \ c_2 = 0, \ c_3 = 0, \ q_{10} = \frac{1}{2}(-1+\sqrt{1+4\Lambda_3}) = \ell, \quad (49)$$

and

$$p_{10} = \sqrt{\Lambda_1} = \sqrt{\frac{2mw}{\hbar^2}}, \quad w = -E, \quad k = 2\ell + 1 \tag{50}$$

Hence we can write the energy eigenvalues and the wave functions as

$$E_{n\ell} = -\frac{me^4}{2\hbar^2 n_0^2}, \quad R_{n\ell} = C\rho^\ell \exp(-\rho) L_{n_0-\ell-1}^{2\ell+1}(2\rho) \tag{51}$$

where C is a constant, $\rho = (\sqrt{2mw/\hbar^2})r$, and $n_0$ is the principal quantum number $n_0 = n - \ell - 1$

Case 5: Pseudoharmonic Potential

Pseudoharmonic potential [39,40] is given by

$$V(r) = V_0 \left(\frac{r}{r_0} - \frac{r_0}{r}\right)^2 \tag{52}$$

and using $s = r^2$ the radial part of the SE takes the form

$$\frac{d^2 R}{ds^2} + \frac{3/2}{s} \frac{dR}{ds} + \frac{1}{s^2}[-\Lambda_1 s^2 + \Lambda_2 s - \Lambda_3]R = 0 \tag{53}$$

where

$$\Lambda_1 = \frac{mV_0}{2\hbar^2}, \quad \Lambda_2 = \frac{m}{2\hbar^2}(E + V_0), \quad \Lambda_3 = \frac{mV_0 r_0^2}{2\hbar^2} + \frac{\ell(\ell+1)}{4}. \tag{54}$$

From Eqs. (1,21,23,26) we obtain

$$c_1 = \frac{2}{3}, \; c_2 = 0, \; c_3 = 0, \; q_{10} = \frac{-1}{4} + \sqrt{\frac{1}{16} + \Lambda_3}, \; p_{10} = \sqrt{\Lambda_1}, \; k = 2\sqrt{\frac{1}{16} + \Lambda_3}. \tag{55}$$

Thus the energy levels and the corresponding wave functions are

$$\varepsilon = [2n + 1 + 2\sqrt{\frac{1}{16} + \beta}]\alpha, \quad \psi = s^{\frac{-1}{4} + \sqrt{\frac{1}{16}+\beta}} \exp(-\alpha s) L_n^{2\sqrt{\frac{1}{16}+\beta}}(2\alpha s) \tag{56}$$

where

$$\varepsilon = \Lambda_2, \quad \alpha^2 = \Lambda_1, \quad \beta = \Lambda_3. \tag{57}$$

Case 6: The Noncentral Potential

The noncetral potential [41-47] is given as

$$V(r,\theta) = \frac{\alpha}{r} + \frac{\beta}{r^2 \sin^2(\theta)} + \gamma \frac{\cos(\theta)}{r^2 \sin^2(\theta)} \tag{58}$$

The radial part of the SE takes the following form

$$\frac{d^2R}{dr^2} + \frac{2}{r}\frac{dR}{dr} + \frac{2m}{r^2}[\frac{2mE}{\hbar^2}r^2 - \frac{2m\alpha}{\hbar^2}r - \frac{2m\lambda}{\hbar^2}]R = 0 \tag{59}$$

where $\lambda$ is a constant. When we define a variable s as s=r this equation becomes

$$\frac{d^2R}{ds^2} + \frac{2}{s}\frac{dR}{ds} + \frac{2m}{s^2}[-\Lambda_1 s^2 + \Lambda_2 s - \Lambda_3]R = 0 \tag{60}$$

where

$$\Lambda_1 = -\frac{2mE}{\hbar^2} = -\varepsilon^2, \quad \Lambda_2 = -\frac{2m\alpha}{\hbar^2}, \quad \Lambda_3 = \frac{2m\lambda}{\hbar^2}, \quad \lambda = \frac{\hbar^2 \ell(\ell+1)}{2m} \tag{61}$$

We use Eqs.(1,18,21,23,26) and obtain

$$c_1 = 2, \; c_2 = 0, \; c_3 = 0, \; q_{10} = \frac{-1}{2} + \sqrt{\frac{1}{4} + \Lambda_3}, \; p_{10} = \sqrt{\Lambda_1}, \; k = 2\sqrt{\frac{1}{16} + \Lambda_3}. \tag{62}$$

The energy eigenvalues are

$$E_n = -\frac{2m}{\hbar^2}\frac{\alpha^2}{[2n+1+\sqrt{1+4\Lambda_3}]^2} = -\frac{2m}{\hbar^2}\frac{\alpha^2}{4[n+\ell+1]^2} \tag{63}$$

And the corresponding wave functions are

$$\psi_{n\ell} = r^\ell \exp(-\varepsilon r) L_n^{2\ell+1}(2\varepsilon r) \tag{64}$$

B) nonzero $c_3$ cases.

Case 7: Deformed Rosen-Morse Potential

The Deformed Rosen-Morse Potential [48-53] has the following form

$$V(x) = \frac{V_1}{[1+\eta\exp(-2ax)]} - \frac{V_2\eta\exp(-2ax)}{[1+\eta\exp(2ax)]^2}. \tag{65}$$

The SE becomes

$$\frac{d^2\psi}{ds^2} + \frac{(1-\eta s)}{s(1-\eta s)}\frac{d\psi}{ds} + \frac{1}{[s(1-\eta s)]^2}[-\Lambda_1 s^2 + \Lambda_2 s - \Lambda_3]\psi = 0 \tag{66}$$

where

$$\Lambda_1 = \varepsilon\eta^2, \quad \Lambda_2 = 2\varepsilon\eta + \kappa\eta - \gamma, \quad \Lambda_3 = \varepsilon + \kappa. \tag{67}$$

$$\varepsilon = -\frac{mE}{2\hbar^2\alpha^2}, \quad \kappa = \frac{mV_1}{2\hbar^2\alpha^2}, \quad \gamma = \frac{mV_2\eta}{2\hbar^2\alpha^2}$$

Using Eqs. (1,4b,2,7,9,) we obtain the following parameters

$$c_1 = 1, \; c_2 = -\eta, \; c_3 = -\eta, \; q_0 = \sqrt{\Lambda_3} = \sqrt{\varepsilon+\kappa}, \; p_0 = \frac{-1}{2}(1+\sqrt{1+4\frac{\gamma}{\eta}}) \tag{68}$$

and

$$\alpha = 2\sqrt{\varepsilon+\beta}, \quad \beta = \sqrt{1+4\frac{\gamma}{\eta}}. \tag{69}$$

Let us define $x = (p_0 - q_0)$ and insert the values of $c_i$ into Eq. (13) it becomes $x^2 - 2nx + n^2 - \varepsilon = 0$. We use the solution $x = n + \sqrt{\varepsilon}$ and write $p_0 - q_0 = n + \sqrt{\varepsilon}$ as

$$n + \sqrt{\varepsilon} = -\frac{1}{2}(1+\sqrt{1+4\frac{\gamma}{\eta}}) - \sqrt{\varepsilon+\kappa} = 0. \tag{70}$$

Solving this equation for $\varepsilon$ we obtain

$$\varepsilon - \frac{\kappa}{2} + \frac{1}{16}[2n+1+\sqrt{1+4\frac{\gamma}{\eta}}] + [\frac{\kappa}{2n+1+\sqrt{1+4\frac{\gamma}{\eta}}}]^2 \qquad (71)$$

and for the corresponding wave functions are

$$\psi = (1-qs)^{\frac{1}{2}(1+\sqrt{1+4\frac{\gamma}{\eta}})} s^{\sqrt{\varepsilon+\kappa}} P_n^{(2\sqrt{\varepsilon+\kappa},\sqrt{1+4\frac{\gamma}{\eta}})}(1-2qs). \qquad (72)$$

Case 8: Woods-Saxon Potential

The Generalized Woods-Saxon potential [54-60] is

$$V(x) = \frac{-V_1}{1+\exp(ax)} - V_2 \frac{\exp(ax)}{[1+\exp(ax)]^2} \qquad (73)$$

and defining $s = 1/[1+\exp(ax)]$ the SE can be written as

$$\frac{d^2\psi}{ds^2} + \frac{1-2s}{s(1-s)}\frac{d\psi}{ds} + \frac{1}{[s(1-s)]^2}[-\Lambda_1 s^2 + \Lambda_2 s - \Lambda_3]\psi = 0 \qquad (74)$$

where

$$\Lambda_1 = \frac{2ma^2 V_2}{\hbar^2}, \quad \Lambda_2 = \frac{2ma^2}{\hbar^2}(V_2+V_1), \quad \Lambda_3 = -\frac{2ma^2 E}{\hbar^2}. \qquad (75)$$

Thus the parameters of Eq. (1) are $c_1 = 1$, $c_2 = -2$, $c_3 = -1$. We calculate $q_0, p_0$ by means of Eq. (7) and Eq. (9) and find

$$q_0 = \sqrt{\varepsilon}, \quad p_0 = \sqrt{\varepsilon-a} \qquad (76)$$

where we $\varepsilon$ is used instead of $\Lambda_3$. The energy levels are obtained with Eq.(13) as

$$x^2 + (2n+1)x + n(n+1) - \Lambda_1 = 0 \tag{77}$$

where $x = \sqrt{\varepsilon} - \sqrt{\varepsilon - a}$. The solution of this quadratic equation gives

$$x = (1/2)[-(2n+1) + \sqrt{1+4\gamma}] \tag{78}$$

where we have used $\gamma$ instead of $\Lambda_1$. Solving Eq. (40) for $\varepsilon$ we find

$$\varepsilon = \frac{1}{16}[-(2n+1) + \sqrt{1+4\gamma}]^2 + \frac{a}{2} + \frac{a^2}{[-2n+1) + \sqrt{1+4\gamma}]^2} \tag{79}$$

The corresponding wave functions can be written from Eq. (2) and Eq. (4b) as

$$\psi = s^{\sqrt{\varepsilon}}(1-s)^{-\sqrt{\varepsilon-a}} P_n^{(2\sqrt{\varepsilon},-2\sqrt{\varepsilon-a})}(1-2s). \tag{80}$$

Case 9: The Pöschl-Teller Potential

The Pöschl-Teller potential[61-66] is given by

$$V(x) = -4V_0 \frac{\exp(-2ax)}{[1+\exp(-2ax)]^2}. \tag{81}$$

and the SE takes the form

$$\frac{d^2\psi}{ds^2} + \frac{(1-\eta s)}{s(1-\eta s)} \frac{d\psi}{ds} + \frac{1}{[s(1-\eta s)]^2}[-\Lambda_1 s^2 + \Lambda_2 s - \Lambda_3]\psi = 0 \tag{82}$$

where $\Lambda_1 = -\varepsilon^2 q^2$, $\Lambda_2 = 2\varepsilon^2 q - \beta^2$, $\Lambda_3 = \varepsilon^2$, $\varepsilon^2 = \frac{2ma^2}{\hbar^2}E$

By using Eqs.(1,4b,2,7,9) we find

$$c_1 = 1, \ c_2 = -\eta, \ c = -\eta, \ q_0 = \varepsilon, \ p_0 = -\frac{1}{2}(1 + \sqrt{1 + 4\frac{\beta^2}{\eta}}). \tag{83}$$

and

$$\alpha = 2\varepsilon, \ \beta = 2\sqrt{1 + 4\frac{\beta^2}{\eta}}. \tag{84}$$

We write Eq. (13) as $x^2 - 2x + n^2 - \varepsilon^2 = 0$ where $x = p_0 - q_0$. Choose the root $x = n + \varepsilon$ and replace this in the definition as

$$n + \varepsilon = -\frac{1}{2}(1 + \sqrt{1 + 4\frac{\beta^2}{\eta}}) - \varepsilon. \tag{85}$$

Thus we get the following energy values and the wave functions

]
$$\varepsilon = -\frac{1}{4}(2n + 1 + \sqrt{1 + 4\frac{\beta^2}{\eta}}), \quad \psi = (1 - qs)^{\frac{1}{2}(1 + \sqrt{1 + 4\frac{\beta^2}{\eta}})} s^{2\varepsilon} P_n^{(2\varepsilon, 2\sqrt{1 + 4\frac{\beta^2}{\eta}})} \tag{86}$$

## 4. Conclusions

We can conclude by saying that, for several physical potentials the Schrödinger can be transformed into a second order differential equation of a certain form. We have shown that differential equations of that form can be solved analytically. Starting from the parametric form of the differential equation we have developed a formulation for the possible physically acceptable solutions. To show that the present method is an efficient and practical method we have applied it to several potentials.

## Appendix A

The polynomial method is a useful method for the solution of the Schrodinger equation. A power series expansion of the wave function is inserted into the wave equation and the coefficients are determined. We want to apply this method to Eq. (4a). Let us write that equation again

$$(1-z^2)^2 \frac{d^2 y}{dz^2} + (1-z^2)[\beta - \alpha + (\alpha + \beta + 2)z]\frac{dy}{dz} + R(z)y = 0. \tag{A1}$$

In this equation if z is set equal to one all the terms except the last one vanish. Since $y$ is assumed to be nonzero at this point, it follows that R must vanish. That means $r_1 + r_2 + r_3 = 0$ must be satisfied. A similar analysis at z=-1 gives $r_1 - r_2 + r_3 = 0$. Combining these two equations we get $r_2 = 0$ and $r_1 = -r_3$. Since we have two arbitrary constants in our formalism it may be possible to satisfy these restrictions. Inserting these into Eq. (4a) we obtain

$$(1-z^2)\frac{d^2 y}{dz^2} + [\beta - \alpha + (\alpha + \beta + 2)z]\frac{dy}{dz} + r_3 y = 0. \tag{A2}$$

Let us represent $y$ as $\sum_v a_v z^v$ and insert this into Eq. (A2). Since the right hand-side of the equation is zero the coefficients each power of z most vanish. When we consider the coefficient of $z^v$ for large enough values of $v$, we get an asymptotic relation $a_{v+1}/a_v \cong 1$. Hence this infinite series will not lead to an acceptable wave function when inserted into Eq.(2). Thus the series must break off after a finite number of terms. If we assume that it breaks off at $v = n$ then the solution will be represented as $\sum_v^n a_v z^v$. Inserting this into Eq.(A2) we find the coefficient of $z^n$ as $[-n(n-1) - (\alpha + \beta + 2)n + r_3]a_n$. Since the coefficient of $z^n$ must vanish and $a_n$ is not zero we obtain $r_3 = n(n + \alpha + \beta + 1)$.

For the $c_3 = 0$ case we try the same method for Eq.(17). Using $\gamma_3 = 0$ in Eq.(17) we write that equation as

$$z\frac{d^2 y_1}{dz^2} + (k+1-z)\frac{dy_1}{dz} + (\gamma_1 z + \gamma_2)y_1 = 0 . \tag{A3}$$

Representing $y_1$ as $\sum_v a_v z^v$ and inserting it into Eq.(A3) we found that for large values of $v$ the coefficients will have an asymptotic relation $a_{v+1}/a_v \cong 1/v$. We note that the successive coefficients in the expansion of $e^z$ have this asymptotic behavior. So the asymptotic behavior of $y_1$ can be written as $\exp(z)$ or $\exp[(2p_1 - c_2)s]$. This will be multiplied by the factor $\exp(-p_1 s)$ in Eq.(15). Thus the wave function will have a factor $\exp[(p_1 - c_2)s]$ and $p_1 - c_2$ is positive. This can be seen from Eq.(23). Therefore the series must break off. When we replace $y_1$ by $\sum_v^n a_v z^v$ in Eq.(A3) it gives the following equation

$$z^{n+1}(\gamma_1 a_n) + z^n(-na_n + \gamma_2 a_n) + z^{n-1}(...) + ...... = 0 \tag{A4}$$

From this we obtain $\gamma_1 = 0$ and $\gamma_2 = n$. These values are used in the text.